\documentclass[preprint,12pt]{elsarticle}

\usepackage{amssymb}
\usepackage{amsmath}
\usepackage{url}
\usepackage{hyperref}
\usepackage{xcolor}
\usepackage{braket}
\usepackage{xspace}
\usepackage{pgf,tikz}
\usepackage{pifont}
\usepackage{braket}

\usepackage{listings}
\usepackage{cleveref}
\crefname{appendix}{}{}
\usepackage{easy-todo}

\usetikzlibrary{calc}
\usetikzlibrary{arrows}
\usetikzlibrary{decorations.markings,decorations.text,calc,arrows.meta,positioning}

\tikzset{
  state/.style={circle,draw,minimum size=6ex},
  arrow/.style={-latex, shorten >=1ex, shorten <=1ex}}

\biboptions{numbers,sort&compress}

\newcounter{bla}

\journal{Computer Physics Communications}

\def\Caravel{{\textsc{Caravel}}}
\def\Cpp{\textsc{C\nolinebreak[4]\hspace{-.05em}\raisebox{.4ex}{\tiny\bf ++}}}
\def\Cppstd{\textsc{C\nolinebreak[4]\hspace{-.05em}\raisebox{.4ex}{\tiny\bf ++}17}}

\definecolor{myblue}{rgb}{0,0.35,0.638}
\definecolor{myred}{RGB}{193,0,13}
\definecolor{mygrey}{RGB}{124,125,126}
\definecolor{mybrown}{RGB}{180, 90, 80}

\begin{document}

\begin{frontmatter}



\title{
\hbox{\rm\small
CP3-20-44, FR-PHENO-2020-009, MPP-2020-170$\null\hskip 2.3cm \null$
\break}
\hbox{$\null$\break}
\Caravel{}: A \texorpdfstring{\Cpp{}}{C++} Framework for the Computation of Multi-Loop
Amplitudes with Numerical Unitarity}


\author[a]{S.~Abreu}
\author[b]{J.~Dormans}
\author[c]{F.~Febres~Cordero\corref{author}}
\author[b]{H.~Ita}
\author[c]{M.~Kraus}
\author[d]{B.~Page\corref{author}}
\author[b]{E.~Pascual}
\author[b]{M.~S.~Ruf}
\author[e]{V.~Sotnikov\corref{author}}

\cortext[author] {Corresponding authors.\\\textit{E-mail addresses:}
ffebres@hep.fsu.edu, page@ipht.fr, sotnikov@mpp.mpg.de}

\address[a]{Center for Cosmology, Particle Physics and Phenomenology (CP3),
Universit\'{e} Catholique de Louvain, 1348 Louvain-La-Neuve, Belgium}
\address[b]{Physikalisches Institut, Albert-Ludwigs-Universit\"at Freiburg,
Hermann-Herder-Str.~3, D-79104 Freiburg, Germany}
\address[c]{Physics Department, Florida State University, 77 Chieftan Way,
Tallahassee, FL 32306, U.S.A.}
\address[d]{Institut de Physique Th\'eorique, CEA, CNRS, Universit\'e
Paris-Saclay, F-91191 Gif-sur-Yvette cedex, France}
\address[e]{Max-Planck-Institut f\"ur Physik, Werner-Heisenberg-Institut, 80805
M\"unchen, Germany}

\begin{abstract}
We present the first public version of \Caravel{}, a \Cppstd{} framework for
the computation of multi-loop scattering amplitudes in quantum field theory, based on
the numerical unitarity method. \Caravel{} is composed of modules for
the $D$-dimensional decomposition of integrands of scattering amplitudes into master and
surface terms, the computation of tree-level amplitudes in floating point or
finite-field arithmetic, the numerical computation of one- and two-loop
amplitudes in QCD and Einstein gravity, and functional reconstruction tools.
We provide programs that showcase \Caravel{}'s main functionalities and allow
to compute selected one- and two-loop amplitudes.

\end{abstract}


\end{frontmatter}



\vspace{5mm}
{\bf PROGRAM SUMMARY}

\vspace{5mm}
\begin{small}
\noindent
{\em Program Title: } \Caravel{}     \\
{\em Licensing provisions: } GPLv3     \\
{\em Programming language: } \Cpp{}    \\
{\em Required compilers: } \texttt{g++} or \texttt{clang} with \Cppstd{} support    \\
{\em Required external dependencies: } \texttt{Python3}~\cite{1}, \texttt{meson}~\cite{2} \\
{\em Optional external dependencies: }
\texttt{Doxygen}~\cite{3},
\texttt{Eigen}~\cite{4},
\texttt{GiNaC}~\cite{5},
\texttt{GMP}~\cite{6},
\texttt{Lapack}~\cite{7},
\texttt{MPFR}~\cite{8},
\texttt{MPI}~\cite{9}, 
\texttt{PentagonLibrary}~\cite{10,11}, 
 \texttt{QD}~\cite{12}\\
{\em Computer(s) for which the program has been designed:} \\
From personal laptops to supercomputers \\
{\em Has the code been vectorized or parallelized?: }  \\
Yes       \\
{\em Number of processors used: } \\ The program can run single threaded, in a
processor with multi-threading, or in a supercomputer. \vspace{5mm}\\
{\em Supplementary material:} \\
The article, \url{https://gitlab.com/caravel-public/caravel.git}  \\
{\em Nature of problem:} \\
The computation of multi-loop multi-particle scattering amplitudes in quantum field 
theory\\
{\em Solution method:} \\
The multi-loop numerical unitarity method, functional reconstruction algorithms\\
{\em Additional comments including restrictions and unusual features:} \\
 Current version includes tools employed in previous
calculations, with the aim of showcasing details of the algorithms employed.
Computations are organized by provided data files.\\
   \\

\end{small}

\tableofcontents
\section{Introduction}
\label{sec:intro}

The computation of scattering amplitudes in quantum field theory is crucial in
our quest to describe high-energy particle interactions. Indeed,
these objects allow one to make theoretical predictions which can then
be compared with experimental measurements, be it at particle colliders
such as the Large Hadron Collider (LHC), which are testing the Standard Model of particle physics,
or at experiments such as LIGO or VIRGO, which are testing our understanding of
gravity. 
Computing these amplitudes remains a challenge, in particular when they contain many external
particles and when higher-order corrections in their perturbative expansion are
necessary. The former implies a dependence on a large number of physical scales, and
the latter a number of unconstrained \textit{loop momenta} which must be integrated over.
Combined, these two aspects lead to a considerable level of complexity
in the computation of these amplitudes.
This is nevertheless a very timely and important problem to tackle. For example, 
two-loop five-particle amplitudes as well as 
three-loop four-particle amplitudes are already relevant for phenomenological
studies at the LHC, and will become even more so over the next years.

A major obstacle to overcome when evaluating loop amplitudes is the 
complexity of intermediate computational steps. In order to bypass this
issue, it has proven fruitful to consider numerical approaches.
An outstanding example of this is the current possibility to compute general 
one-loop amplitudes, which has been
powered by the introduction of robust numerical techniques. Among many other developments,
these techniques are based on integrand reduction approaches~\cite{delAguila:2004nf,Ossola:2006us}, 
on the one-loop numerical unitarity method~\cite{Ellis:2007br,Giele:2008ve,Berger:2008sj}, 
and on recursive approaches~\cite{Cascioli:2011va,Actis:2012qn}.
More recently, there has been great progress in the numerical computation of two-loop
amplitudes at high multiplicities. By now, all five-parton amplitudes
\cite{Abreu:2017hqn, Badger:2017jhb, Abreu:2018jgq, Badger:2018gip} and the
amplitudes for four partons and a $W$ boson~\cite{Hartanto:2019uvl} have been computed numerically at leading color.
Through the use of finite fields and multivariate functional reconstruction
techniques~\cite{vonManteuffel:2014ixa,Peraro:2016wsq}, the frameworks powering
these numerical computations have furthermore allowed the calculation of the analytic form
of all five-parton leading-color amplitudes~\cite{Badger:2018enw, Abreu:2018zmy,
  Abreu:2019odu}, the five-point all-plus amplitude at full
color~\cite{Badger:2019djh} and the four-graviton amplitude in Einstein
gravity~\cite{Abreu:2020lyk}. In related work, the amplitudes for
three-photon production at the LHC have also been computed~\cite{Chawdhry:2019bji}.

In this article we present the \Caravel{} \Cpp{} framework. It provides an
implementation of many algorithms necessary to perform computations of
multi-loop scattering amplitudes within the multi-loop numerical unitarity
method. This is the first publicly available code of its kind. It is based on the (generalized) unitarity approach, which was first
developed for the analytic computation of one-loop amplitudes~\cite{Bern:1994zx,
Bern:1994cg,Bern:1997sc,Britto:2004nc} and later adapted for numeric calculations~\cite{Ellis:2007br,Giele:2008ve,Berger:2008sj}.
An extension of the method beyond one loop has been developed
recently~\cite{Ita:2015tya,Abreu:2017idw,Abreu:2017xsl}.
In a nutshell, in this framework the amplitude is computed
starting from a parametrization of its integrand. The corresponding free parameters are 
numerically computed at each phase-space point 
by constructing systems of linear equations in which the parameters
are the unknowns and the numerical entries are associated to products of
tree-level amplitudes. With a suitable choice of  integrand
parametrization~\cite{Ita:2015tya}, this directly gives a decomposition of the amplitude
in terms of master integrals. Finally, after inserting the value of the integrals at the
required phase-space point we obtain the value of the amplitude.

The current release of \Caravel{} includes a module for
computing products of tree-level amplitudes in several theories through
off-shell recursion relations~\cite{Berends:1987me}, and tools that
allow the efficient construction and solution of the systems of linear equations that determine the integrand.
Whilst these components work for generic multi-loop amplitudes, other
components such as the construction of the parametrization
are required as input. In this release we showcase the
different available tools by providing a series of example programs.
Discussions regarding full automation are left to future work. 
The example programs give a first-hand account of the procedures employed for the
calculations presented in
refs.~\cite{Abreu:2017xsl,Abreu:2017hqn,Abreu:2018jgq,Abreu:2018zmy,
Abreu:2019odu,Abreu:2020lyk}.

The rest of this article is organized as follows. \Cref{sec:methods}
provides a brief description of our computational methodology,
\cref{sec:structure} gives a description of the organization of the
internal modules in the \Caravel{} framework, \cref{sec:install}
describes the procedure of installation and setup of the libraries. In
\cref{sec:run} we give details on how to run the example programs that we provide and
we conclude in \cref{sec:conclusions}. \cref{sec:helicity_caravel,app:twistor}
contain technical details about our conventions for color-ordered 
helicity amplitudes and phase-space parametrizations.

\section{Computational methodology}
\label{sec:methods}

In this section we briefly review the main features of 
the multi-loop numerical unitarity method. This framework allows for the 
numerical evaluation as well as the analytical reconstruction of multi-loop scattering 
amplitudes. Our approach is generic, in that it facilitates the computation of amplitudes in different
quantum field theories. When dealing with QCD amplitudes, we consider the amplitudes
to be vectors $\mathcal{M}$ in a generic color space. In this section
we will mostly be concerned with the components of these vectors, that is with the
$\mathcal{A}(\sigma)$ defined as
\begin{equation}\label{eq:colorDec}
  \mathcal{M}=\sum_\sigma C(\sigma) \mathcal{A}(\sigma)\,,
\end{equation}
where the $C(\sigma)$ span the relevant color space.
For QCD amplitudes, this is the usual $SU(N_c)$ color space, which can be
specialized to $N_c=3$. For pure gravity processes,
no color is present and in eq.~\eqref{eq:colorDec} there is
a single $\sigma$ with $C(\sigma)=1$. In the remaining of this section, we will
always discuss the $\mathcal{A}(\sigma)$ defined in eq.~\eqref{eq:colorDec}, which for brevity
we will call amplitudes, and for simplicity we will drop the $\sigma$ dependence.
In later sections, when discussing the calculation of amplitudes in specific theories,
we will be more explicit about the $C(\sigma)$.
As is standard in perturbative quantum field theory, we will consider the expansion 
of $\mathcal{A}$ around a small coupling constant, which is associated
to an expansion in the number of loops of the contributing Feynman integrals.
We refer the reader to \ref{sec:helicity_caravel} for more details on the definition
of the objects we compute.
For more details on the techniques outlined in this section we refer to previous 
publications~\cite{Abreu:2017xsl,Abreu:2017hqn,Abreu:2018jgq,Abreu:2018zmy,
Abreu:2019odu,Abreu:2020lyk}.

\subsection{Integrand parametrization}

In full generality, an $L$-loop amplitude $\mathcal{A}^{(L)}$ can be decomposed 
as a linear combination of master integrals according to
\begin{align}
  \mathcal{A}^{(L)} &=  \sum_{\Gamma \in \Delta} \sum_{i \in M_{\Gamma}}
         c_{\Gamma, i}\ \mathcal{I}_{\Gamma, i}\ .
  \label{eq:AmpAnsatz}
\end{align}
Here, $\Gamma$ defines a propagator structure associated with the amplitude,
and we will often refer to it as a \emph{diagram} (indeed, they
are in one-to-one correspondence with scalar Feynman integrals).
The set $\Delta$ contains all propagator
structures $\Gamma$, and can be organized hierarchically according to whether a
propagator structure $\Gamma_1\in\Delta$ can be obtained from another
$\Gamma_2\in\Delta$ by removing some propagators in the latter.
In this case we write $\Gamma_1<\Gamma_2$. 
The set $M_\Gamma$ denotes the full set of master integrals associated to $\Gamma$.
Each master integral $\mathcal{I}_{\Gamma, i}$
is usually expressed as a Laurent series in the dimensional-regularization 
parameter $\epsilon = (4-D)/2$. The coefficients in the Laurent expansion
involve multi-valued functions with non-trivial branch-cut structures,
such as multiple polylogarithms. The coefficients $c_{\Gamma, i}$ 
are algebraic functions of the kinematic invariants and
rational functions of $\epsilon$.

At its core, the unitarity method is a way to compute the integrand of an 
amplitude. We therefore start with a parametrization of the integrand $\mathcal{A}^{(L)}(\ell_l)$, 
where $\ell_l$ represents the set of $L$ loop momenta, of the form
\begin{align}
  \mathcal{A}^{(L)}(\ell_l) &=  \sum_{\Gamma \in \Delta} \sum_{k \in Q_{\Gamma}}
  c_{\Gamma, k} \frac{m_{\Gamma, k}(\ell_l)}{\prod_{j \in P_{\Gamma}} 
  \rho_j(\ell_l)}\ .
  \label{eq:GenericIntegrandAnsatz}
\end{align}
The multiset $P_\Gamma$ labels all inverse propagators $\rho_j$ in the diagram $\Gamma$, 
and the basis of numerators  $\mathcal{Q}_\Gamma=\{m_{\Gamma, k}(\ell_l)|k\in Q_\Gamma\}$ 
parametrizes every possible integrand up to the maximum allowed power of loop momenta, 
which is theory specific. 

The numerator basis $\mathcal{Q}_\Gamma$ is not unique. 
Let us highlight two natural classes of bases.
First, the simplest choice is what we call a \textit{tensor basis}, denoted by
$\mathcal{T}_\Gamma$. It can be built out of independent monomials in a set of 
variables $\alpha_{j}$, 
which parametrize the loop momenta $\ell_l(\vec\alpha)$~\cite{Mastrolia:2011pr,
Badger:2012dp,Zhang:2012ce,Mastrolia:2012an}.
This type of basis is straightforward to build for generic $\Gamma\in\Delta$.
However, with this choice, the relation between eqs.~\eqref{eq:AmpAnsatz}
and \eqref{eq:GenericIntegrandAnsatz} is not explicit, as it would require solving
large systems of integration-by-part (IBP) relations 
(see e.g.\ \cite{Chetyrkin:1981qh,Laporta:2001dd,Anastasiou:2004vj,
Studerus:2009ye,vonManteuffel:2012np,Lee:2013mka,
Smirnov:2013dia,vonManteuffel:2014ixa,Smirnov:2014hma,Maierhoefer:2017hyi}).
A second class of bases is what we call a \textit{master-surface basis}
$\mathcal{M}_\Gamma$, and it is crucial to our multi-loop numerical unitarity
method. It was observed in ref.~\cite{Ita:2015tya} that one can parametrize the
integrand of a multi-loop amplitude by functions
$\mathcal{M}_\Gamma=\{m_{\Gamma,i}(\ell_l)|i\in M_\Gamma\cup S_\Gamma\}$ such 
that the associated integrands either integrate to zero or correspond to the integrand of
a master integral:
\begin{equation} 
\int \left( \prod_{j=1}^L\frac{d^D\ell_j}{(2\pi)^D}\right)\,
\frac{m_{\Gamma,i}(\ell_l)}{\prod_{k\in P_\Gamma} \rho_k(\ell_l)} = \left\{
    \begin{array}{cc} \mathcal{I}_{\Gamma,i} &  \mbox{for}\quad i\in
M_\Gamma\,, \\ 0&  \mbox{for}\quad i\in S_\Gamma\,. \end{array}\right.
\end{equation}
The numerators $m_{\Gamma,i}(\ell_l)$ with $i\in M_\Gamma$ are called 
\textit{master integrands} and the ones with $i\in S_\Gamma$ \textit{surface terms}.  
A master-surface basis of functions thus makes the relation between 
eqs.~\eqref{eq:AmpAnsatz} and \eqref{eq:GenericIntegrandAnsatz} explicit.
As discussed in~\cite{Ita:2015tya,Abreu:2017idw,Abreu:2017xsl} the construction of
master-surface bases of integrands can be efficiently performed by using
unitarity-compatible IBP relations~\cite{Gluza:2010ws,Schabinger:2011dz}, 
employing computational algebraic geometry techniques.

\subsection{Integrand factorization and cut equations}
\label{sec:cut_equations}

In order to compute the coefficients $c_{\Gamma,k}$ in
\cref{eq:GenericIntegrandAnsatz} we exploit the factorization properties of
multi-loop integrands for \textit{on-shell} configurations $\ell_l^\Gamma$ 
of the loop momenta. For a given propagator structure $P_\Gamma$, these are defined by
\begin{equation}
 \rho_j(\ell_l^\Gamma) = 0\,\,\, 
 \text{iff}\,\,\, j \in P_\Gamma\,.
 \label{eqn:onshell}
\end{equation}
In most cases,
this does not fix the $\ell_l^\Gamma$ completely, and there is some residual
degree of freedom. In this limit, the leading pole of
\cref{eq:GenericIntegrandAnsatz} is given by
\begin{align}
  \sum_{\text{states}} \prod_{i \in T_{\Gamma}} \mathcal{A}_i^{\text{tree}} 
  (\ell_l^{\Gamma}) &=
  \sum_{\substack{\Gamma' \geq \Gamma \\ k \in Q_{\Gamma'}}}
  \frac{c_{\Gamma',k} m_{\Gamma', k}(\ell_l^{\Gamma})}{
        \prod_{j \in (P_{\Gamma'} / P_{\Gamma})} \rho_j (\ell_l^{\Gamma})
       }\ ,
  \label{eq:cut_equation}
\end{align}
where $T_\Gamma$ represents the set of all tree-level
amplitudes corresponding to the vertices of the diagram $\Gamma$,
and the state sum runs over all $D_s$-dimensional particle states that can
appear in the loop. On the right-hand side the sum runs over all
propagator structures $\Gamma'$ with equal or more propagators than $\Gamma$ (hence
$P_\Gamma\subseteq P_{\Gamma'}$). 
Eq.~\eqref{eq:cut_equation} is a so-called \textit{cut equation}.

The cut equations allow one to numerically compute the coefficients $c_{\Gamma,k}$ in
\cref{eq:GenericIntegrandAnsatz} by sampling a sufficient
set of on-shell momenta $\ell_l^\Gamma$ and solving the resulting system of linear
equations.
Importantly, some of these coefficients may be identically zero for all
phase-space points. To account for this, we identify zero coefficients during a
``warm-up run'' on a single phase-space point, and remove the corresponding terms
from the ansatz for all subsequent evaluations.
In order to construct this system of equations we must
have an efficient way to evaluate the tree amplitudes on the left-hand
side of the cut equations. This is achieved with an implementation of the Berends-Giele
off-shell recursion relations~\cite{Berends:1987me}, which allows one to
recursively compute tree amplitudes with an arbitrary number of legs, and where the particles
have $D_s$-dimensional states.
Being a very general approach, it provides a straightforward way to add new types of fields.
Beyond one-loop we also need to consider subleading singular contributions for certain
propagator structures for which no generic integrand factorization is known (at two loops, 
this happens for propagator structures where the same propagator appears twice). 
To address these cases, we employ cut equations with fewer on-shell constraints and then 
solve for the corresponding coefficients \cite{Abreu:2017idw}.

Another important aspect in sampling the cut equations is the construction
of the on-shell momenta $\ell^\Gamma$. These configurations of loop
momenta are constructed by solving the quadratic equations in
eq.~\eqref{eqn:onshell}, and depending on which number field we use they
might not have solutions. To be more precise, we will often do calculations
in a field $\mathbb{F}$ that is not algebraically closed, such as the field of rational
numbers or a finite field (see section \ref{sec:exact} below),
which makes the construction of on-shell momenta with components in $\mathbb{F}$
a non-trivial problem. However, it turns out that the square roots originating from the 
solutions of \cref{eqn:onshell} are only present at intermediate steps of the 
calculation, and any $D$-dimensional Lorentz-invariant scalar product of the 
loop momenta lives in $\mathbb{F}$.
In particular, the product of trees in \cref{eq:cut_equation} is free of square roots and 
representable in $\mathbb{F}$. Therefore, we (temporarily) employ an 
algebraic extension of $\mathbb{F}$ for evaluating off-shell currents 
contributing to the left-hand side of \cref{eq:cut_equation}.
We refer to refs.~\cite{Abreu:2018jgq, Abreu:2019odu,Sotnikov:2019onv} for 
details. In some cases, e.g.\ for Yang-Mills theory, it is possible to 
avoid the appearance of square roots altogether \cite{Abreu:2018zmy}.
In this case we compute cuts directly in $\mathbb{F}$.

Solving for all coefficients in an amplitude can be efficiently organized in a
block-triangular way, using the hierarchical structure of the set $\Delta$.
In two-loop five-particle QCD amplitudes each block of equations 
can have up to a few hundreds of unknowns~\cite{Abreu:2017hqn,Abreu:2018jgq},
while in Einstein gravity this number is typically an order of magnitude larger. 
We solve these equations by employing standard linear algebra techniques, such as PLU
or QR factorization. Through this procedure, we can compute the coefficients
$c_{\Gamma,i}$ in eq.~\eqref{eq:AmpAnsatz} at a numerical phase-space point, 
and for numerical values of $D$ and $D_s$.

\subsection{Special functions}
\label{sec:specialFunctions}

As stated below eq.~\eqref{eq:AmpAnsatz}, the master integrals
have a Laurent expansion around $\epsilon=0$, whose coefficients can be
written as linear combinations of multivalued special functions.
For all the cases currently implemented in \Caravel{}, the special
functions are (linear combinations of) multiple polylogarithms.
Using modern mathematical techniques \cite{Goncharov:2010jf,Duhr:2012fh}, we can find a basis $B$
for this space of functions, and find an alternative decomposition
of the amplitude in terms of the elements $h_i\in B$. That is, up to a given
order $k$ in the $\epsilon$ expansion we can write
\begin{equation}\label{eq:specialFuncDec}
  \mathcal{A}^{(L)} =  \sum_{j = - 2 L}^{k} \sum_{i \in B} r_{i,j} \,h_i\, \epsilon^{j}  + \mathcal{O}(\epsilon^{k+1}),
\end{equation}
where the functions $r_{i,j}$ do not depend on $\epsilon$.
This decomposition presents a major difference compared to the one
of eq. \eqref{eq:AmpAnsatz}: it allows one to write one- and two-loop amplitudes
as a linear combination of the same basis of functions. In turn, 
this then makes it possible to write quantities derived from amplitudes, such
as finite remainders, in terms of this basis of functions.
This observation was fundamental in reconstructing two-loop 
five-parton QCD amplitudes \cite{Abreu:2018zmy,Abreu:2019odu}, 
using the basis of ref.~\cite{Gehrmann:2018yef},
as the coefficients in a decomposition
of the form of eq.~\eqref{eq:specialFuncDec} are much simpler for
a two-loop finite remainder than for a two-loop amplitude.
\subsection{Analytic structure in the dimensional regulators}
\label{sec:integrands}

The cut equation in eq.~\eqref{eq:cut_equation} depends on dimensional 
regulators, namely on~$D$, the dimension of the loop momenta, and $D_s$, the 
dimension of the states of loop particles. For convenience, we keep these quantities 
separate until the final stages of the computation. 
The coefficient functions $c_{\Gamma,i}$ in \cref{eq:AmpAnsatz} 
are rational functions in $D$ and polynomials in $D_s$ (in some very special cases, 
this dependence can also be rational, see e.g.~\cite{Abreu:2020lyk}).

We evaluate the products of tree-level amplitudes in \cref{eq:cut_equation} 
in integer $D_s$ dimensions.
To reconstruct the analytic $D_s$ dependence we can employ two different approaches.
In the first approach, known as dimensional reconstruction~\cite{Giele:2008ve,
Ellis:2008ir,Boughezal:2011br,Abreu:2017xsl,Abreu:2017hqn,
Abreu:2018jgq}, we extract the polynomial dependence by evaluating the 
tree-level amplitudes  for various integer dimensions 
$D_s$ and fit the resulting coefficients of the $D_s$ polynomial.
This procedure is conceptually straightforward. However, its numerical 
inefficiency can become a bottleneck for amplitudes with 
external fermions due to the exponential scaling of the dimension of spinor representations 
with $D_s$~\cite{Anger:2018ove,Abreu:2018jgq}. To address this issue, we 
employ a second approach, which bypasses the dimensional reconstruction
method and provides a diagrammatic representation of the coefficients of the $D_s$ polynomial
\cite{Anger:2018ove,Abreu:2019odu,Sotnikov:2019onv}.
As an example, this strategy reduces the evaluation time of two-loop five-parton 
amplitudes with two external fermion lines by about two orders of magnitude. 

Regarding the dependence on $D$, we sample at a sufficient number of values of $D$ in 
order to reconstruct the rational dependence of each master-integral 
coefficient using Thiele's formula~\cite{Peraro:2016wsq, abramowitz1964handbook}.
This procedure can be computationally intensive. 
We note nevertheless that the $D$-dependence in the denominator of the
coefficients is rather simple and independent of the phase-space point.
When evaluating the same amplitude over several phase-space points, we thus
usually perform a warm-up evaluation, dedicated to determining
this dependence for each $c_{\Gamma,k}$ in eq.~\eqref{eq:GenericIntegrandAnsatz}.

\subsection{Finite fields and functional reconstruction}\label{sec:exact}

In ref.~\cite{vonManteuffel:2014ixa} it was shown that finite fields can be
applied to the Laporta algorithm~\cite{Laporta:2001dd} for the IBP reduction of
multi-loop Feynman integrals. The authors were able to not only efficiently
perform numerical IBP reductions of integrals, but also to reconstruct the
analytic rational dependence of the master integral coefficients in the
dimensional regularization parameter $\epsilon$ from those numerical reductions.
It was later shown in ref.~\cite{Peraro:2016wsq} that through a recursive
approach one could more generically reconstruct multivariate rational functions.
In the same paper, this idea was applied to the computation of scattering
amplitudes in generalized unitarity methods.
In \Caravel{}, we apply the reconstruction approach to rational coefficient
functions in the numerical unitarity method. To do this, we evaluate the
amplitudes at rational phase-space points and perform all calculations in a
finite field, obtaining exact
numerical values for the coefficients. These 
evaluations can then be used to reconstruct the analytic dependence on the
kinematic variables which parametrize the appropriate Lorentz-invariant phase space 
associated to the amplitude. \Caravel{} contains all the functionalities for the
numerical evaluation of scattering amplitudes in a finite field, as well as for
the reconstruction of generic rational
functions~\cite{Abreu:2018zmy,Abreu:2019odu}.\footnote{Two publicly available
packages for functional reconstruction are
\texttt{Firefly}~\cite{Klappert:2019emp, Klappert:2020aqs} and
\texttt{FiniteFlow}~\cite{Peraro:2019svx}.} These tools have been fundamental
for the computation of the planar two-loop five-parton
amplitudes~\cite{Abreu:2017hqn,Abreu:2018jgq,Abreu:2018zmy,Abreu:2019odu}, as
well as the two-loop four-graviton amplitudes~\cite{Abreu:2020lyk}.

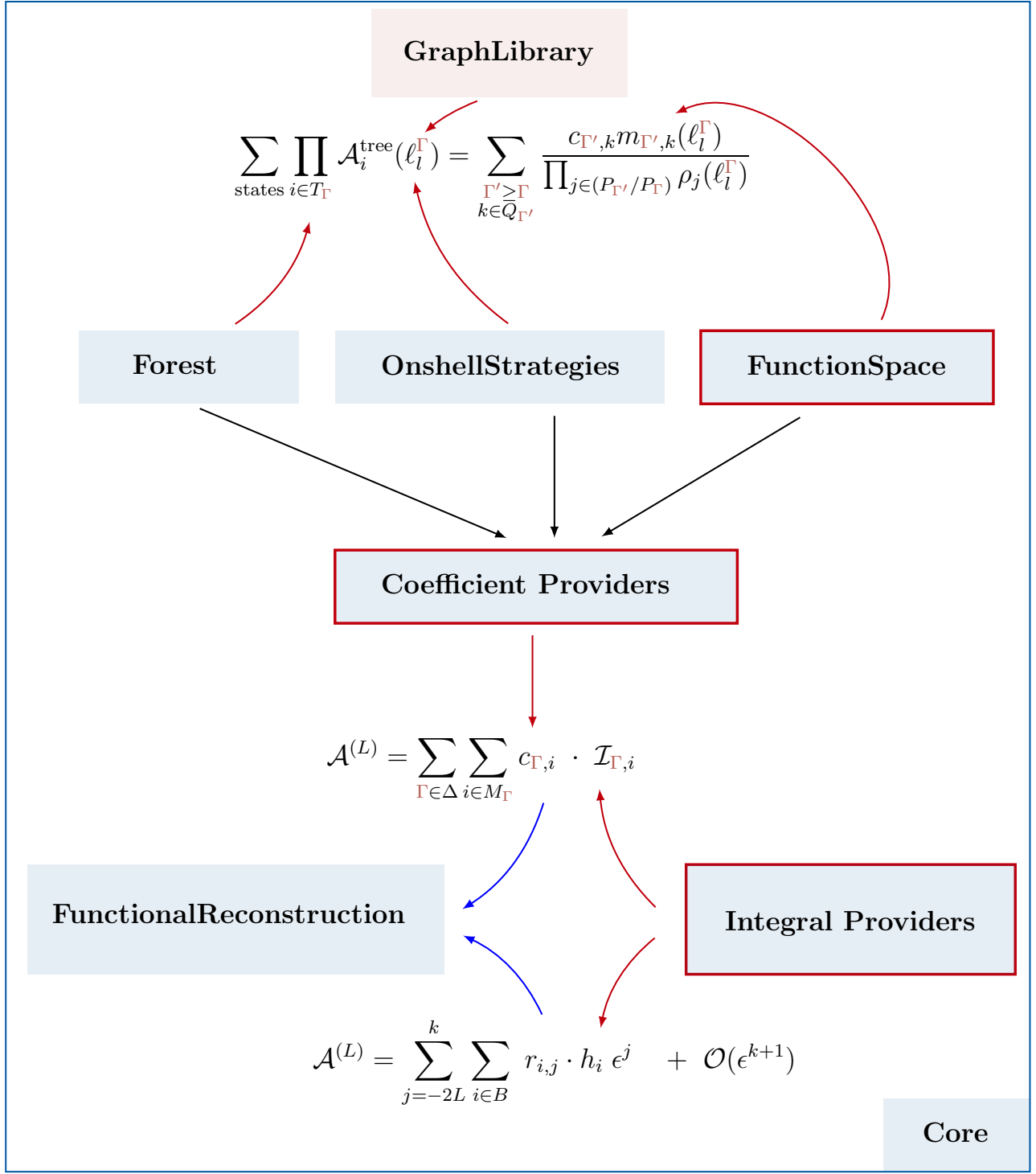
\begin{figure}
      \begin{tikzpicture}[scale=1.1, transform canvas={scale=1.1}]
          \draw[color=myblue, line width=0.7] (-1,1.5) -- (-1,-14.5) -- (13,-14.5) -- (13,1.5) -- cycle;
          \draw (2,0) node[anchor=north west] {$ \displaystyle 
            \sum_{\text{states}} \prod_{i \in T_{\textcolor{mybrown}{\Gamma}}} \mathcal{A}_i^{\text{tree}} 
            (\ell_l^{\textcolor{mybrown}{\Gamma}}) =
            \sum_{\substack{\textcolor{mybrown}{\Gamma'} \geq \textcolor{mybrown}{\Gamma} \\ k \in Q_{\textcolor{mybrown}{\Gamma'}}}}
            \frac{c_{\textcolor{mybrown}{\Gamma'},k} m_{\textcolor{mybrown}{\Gamma'}, k}(\ell_l^{\textcolor{mybrown}{\Gamma}})}{
                  \prod_{j \in (P_{\textcolor{mybrown}{\Gamma'}} / P_{\textcolor{mybrown}{\Gamma}})} \rho_j (\ell_l^{\textcolor{mybrown}{\Gamma}})
                 } $};
          \fill[color=myblue,fill=mybrown, opacity=0.1] (4,0.2) -- (4,1.4) -- (7.5,1.4) -- (7.5,0.2) -- cycle;
          \draw (4.3,1.1) node[anchor=north west] {$\textbf{GraphLibrary}$};
          \draw [arrow, bend angle=10, bend right, line width=0.7, color=myred]  (5.6,0.2) to (4.6, -0.4);
          \draw [arrow, bend angle=20, bend right, line width=0.7, color=myred]  (2,-3) to (3.2,-1.35);
          \draw [arrow, bend angle=20, bend left, line width=0.7, color=myred]  (6,-3) to (4.55,-0.8);
          \draw [arrow, bend angle=70, bend right, line width=0.7, color=myred]  (10.9,-3) to (8.0,-0.1);
          \draw [arrow, line width=0.7] (1.5,-4) to (6.0,-5.9) ;
          \draw [arrow, line width=0.7]  (6.5,-4) to (6.5,-6) ;
          \draw [arrow, line width=0.7]  (10,-4.1) to (7.0,-5.9);
          \draw [arrow, line width=0.7, color=myred]  (6.2,-7) to (6.2,-8.6);
          \draw (5.5, -9) node[] {$ \displaystyle \mathcal{A}^{(L)} = \sum_{\textcolor{mybrown}{\Gamma} \in \Delta} \sum_{i \in M_{\textcolor{mybrown}{\Gamma}}} c_{\textcolor{mybrown}{\Gamma},i} ~ \cdot ~ \mathcal{I}_{\textcolor{mybrown}{\Gamma}, i} $};
          
          \draw (6.5, -13) node[] {$ \displaystyle   \mathcal{A}^{(L)} =  \sum_{j = - 2 L}^{k}  \sum_{i \in B} ~ r_{i,j} \cdot h_i\; \epsilon^{j}  \quad+~ \mathcal{O}(\epsilon^{k+1}) $};
          %
          %
          %
          \fill[color=myblue,fill=myblue,fill opacity=0.1] (0,-3) -- (0,-4) -- (3,-4) -- (3,-3) -- cycle;
          \draw (0.6,-3.2) node[anchor=north west] {$\textbf{Forest}$};
          \fill[color=myblue,fill=myblue,fill opacity=0.1] (3.5,-3) -- (3.5,-4) -- (8,-4) -- (8,-3) -- cycle;
          \draw (4,-3.2) node[anchor=north west] {$\textbf{OnshellStrategies}$};
          \fill[color=myblue,fill=myblue,fill opacity=0.1] (8.5,-3) -- (8.5,-4) -- (12.5,-4) -- (12.5,-3) -- cycle;
          \draw[color=myred,line width=1.2] (8.5,-3) -- (8.5,-4) -- (12.5,-4) -- (12.5,-3) -- cycle;
          \draw (9,-3.2) node[anchor=north west] {$\textbf{FunctionSpace}$};
          \fill[color=myblue,fill=myblue,fill opacity=0.1] (3.5,-6) -- (3.5,-7) -- (9.0,-7) -- (9.0,-6) -- cycle;
          \draw[color=myred, line width=1.2] (3.5,-6) -- (3.5,-7) -- (9.,-7) -- (9.0,-6) -- cycle;
          \draw (4.0,-6.2) node[anchor=north west] {$\textbf{Coefficient Providers}$};
          \fill[color=myblue,fill=myblue,fill opacity=0.1] (-0.7,-10.3) -- (-0.7,-11.8) -- (5.,-11.8) -- (5.,-10.3) -- cycle;
          \draw (-0.5,-10.7) node[anchor=north west] {$\textbf{FunctionalReconstruction}$};

          \draw [color=blue, arrow, bend angle=20, bend left, line width=0.7]  (6.4,-9.3) to (5.1,-11);
          \draw [color=blue, arrow, bend angle=20, bend right, line width=0.7]  (6.4,-12.5) to (5.1,-11.2);
          
          \draw[color=myred, line width=1.2] (8.3,-10.3) -- (8.3,-11.8) -- (12.8,-11.8) -- (12.8,-10.3) -- cycle;
          \fill[color=myblue,fill=myblue,fill opacity=0.1] (8.3,-10.3) -- (8.3,-11.8) -- (12.8,-11.8) -- (12.8,-10.3) -- cycle;
          \draw (8.7,-10.8) node[anchor=north west] {$\textbf{Integral Providers}$};
          %
          \draw [arrow, bend angle=20, bend left, line width=0.7, color=myred]  (8.,-11.) to (7.1,-9.1);
          \draw [arrow, bend angle=20, bend right, line width=0.7, color=myred]  (8.,-11.2) to (7.1,-12.7);

          \fill[color=myblue,fill=myblue,fill opacity=0.1] (11,-13.5) -- (11.,-14.5) -- (13,-14.5) -- (13,-13.5) -- cycle;
          \draw (11.4,-13.7) node[anchor=north west] {$\textbf{Core}$};
          %
          
          %
          %
          
      \end{tikzpicture}
  \vspace{17.5cm}
  \caption{
  Illustration of the interplay of the different modules in \Caravel{} in generic calculations
  of scattering amplitudes. A black arrow indicates a dependence,
  a blue arrow means input for a module, and a red arrow
  the capacity of a module to deliver a given component of the calculation.
  Modules surrounded by a red box rely on external input to operate.
  All modules depend on the \textbf{Core} module.
  }
  \label{fig:caravel_interplay}
\end{figure}

\section{Internal modules}
\label{sec:structure}

The \Caravel{} framework is organized in a modular fashion.
The structure of the multi-loop numerical unitarity method 
outlined in the previous section naturally lends itself to this
modular implementation.
In \cref{fig:caravel_interplay} we show how the different
modules of \Caravel{} relate to the main equations of the numerical
unitarity method 
as well as to each other. The red arrows show which module constructs
each of the different components of these equations.
A black arrow from $A$ to $B$ represents a dependence of $B$ on $A$.
The blue arrows highlight an input that a module receives.
The modules highlighted with red boxes receive external input to
operate, either in the form of data files or as machine-generated source code. 

In the following we list the modules of \Caravel{}, such that the reader gets a 
general view of the source code of the library. We do not give details 
on application programming interfaces, as in this release we only include specific
example programs for the user, as presented in \cref{sec:run}. The modules are:

\begin{itemize}
  \item \textbf{Core}: This module implements general tools for debugging, arithmetic,
kinematics, as well as utilities for linear algebra, rational
reconstruction, type traits, and more general operations such as
Laurent expansions. Among the arithmetic
tools, we include interfaces to the \texttt{QD}~\cite{QD} library for double-double and
quad-double precision floating-point, and the 
\texttt{GMP}~\cite{Granlund12} library for arbitrary precision floating-point
as well as arbitrary size rational numbers. Furthermore,
we use an in-house implementation of 32-bit finite fields based on Barrett
reduction \cite{Barrett:1987} which allows finite fields of cardinality in the
range $(2^{30}, 2^{31}-1]$.
The module also provides linear algebra facilities for solving the cut equations.
These include in-house implementations of linear algebra algorithms for finite fields,
as well as interfaces to \texttt{Lapack}~\cite{laug} (for double precision) 
and to \texttt{Eigen}~\cite{Eigen} (for the double and the high-precision 
floating-point numbers provided by \texttt{QD} and \texttt{GMP}).
Furthermore, the
module contains implementations of $D$-dimensional vector and spinor
representations, designed to work with both floating-point and exact number
types. There is also a parser to process headed lists like those employed in
\texttt{Mathematica}~\cite{Mathematica}.
  \item\textbf{GraphLibrary}: A module for
the classification and canonicalization of multi-loop graphs. In \Caravel{},
many objects can naturally be associated to graphs, such as the
propagator structures $\Gamma$ in \cref{eq:AmpAnsatz}.
Graph isomorphisms are identified by building 
a partial order in the representation of the graph (which is ultimately based on 
the standard \Cpp{} function \texttt{std::lexicographical\_compare}).
  \item\textbf{FunctionalReconstruction}: Implementations of algorithms for
analytic reconstruction of univariate and multivariate rational functions from 
exact numerical evaluations (see \cref{sec:exact}). The 
reconstruction algorithms are parallelized using either native \Cpp{} 
threads or using \texttt{MPI}. The latter is useful for use on computer 
clusters. 
  \item\textbf{OnShellStrategies}: Tools to generate
on-shell loop momenta for  one- and
two-loop diagrams, as required for the construction of the linear system of
equations in \cref{eq:cut_equation}.
  \item\textbf{Forest}: Implementation of the Berends-Giele off-shell 
recursion~\cite{Berends:1987me} for the computation of general tree-level 
amplitudes and the products of trees on the left-hand side of 
\cref{eq:cut_equation} in arbitrary $D_s$ dimensions. 
The recursions can be constructed from any given set of Feynman rules\footnote{
  In particular, there is no restriction on the number of particles in vertices.
}
and can be evaluated over an arbitrary numerical type (e.g.\ floating point of 
different sizes, finite fields, etc.). This release of \Caravel{} 
includes the Feynman rules for massless QCD and Einstein gravity.
  \item\textbf{FunctionSpace}: Module for the construction of the integrand ans\"{a}tze
of \cref{eq:GenericIntegrandAnsatz}, both for tensor bases $\mathcal{T}_\Gamma$
and master-surface bases $\mathcal{M}_\Gamma$. The former can be constructed for
an arbitrary two-loop diagram $\Gamma$, while the latter are provided for arbitrary 
one-loop diagrams and only for the two-loop diagrams required for the calculations
in~\cite{Abreu:2017xsl,Abreu:2017hqn,Abreu:2018jgq,Abreu:2018zmy,Abreu:2019odu,
Abreu:2020lyk}.
Master-surface bases have been produced with private computer-algebra programs, and transformed into \Cpp{}
code to be handled by this module.
  \item\textbf{Integral Providers}: Two separate modules are included to
handle analytic expressions of master integrals. 
The main module is the \textbf{IntegralLibrary}, which provides
access to the master integrals associated to four- and five-point one- and
two-loop planar massless master integrals. In the five-point case, integrals are
written in terms of \emph{pentagon functions}~\cite{Gehrmann:2018yef}.
Internally, all master integrals are normalized as
\begin{equation}\label{eq:normInt}
 \mathcal{I}_{\Gamma,i} = \left(\frac{e^{\gamma_E\epsilon}}{i \pi^{D/2}}\right)^L
 \int  \prod_{j=1}^L d^D\ell_j\,
\frac{m_{\Gamma,i}(\ell_l)}{\prod_{k\in P_\Gamma} \rho_k(\ell_l)} 
\end{equation}
where $\gamma_E$ is the Euler-Mascheroni constant.
In the four-point case, master integrals are evaluated with 
\texttt{GiNaC}~\cite{Bauer:2000cp,Vollinga:2004sn,Vollinga:2005pk} and
\texttt{CLN}~\cite{CLN}. In the five-point case, we currently employ a modified version of
the library provided in~\cite{Gehrmann:2018yef}.
All master integrals are stored in a
format that allows on-the-fly expression of an amplitude in terms of a set of
basis functions as described in section \ref{sec:specialFunctions}, 
see eq.~\eqref{eq:specialFuncDec}.
In the current version all master integrals are implemented in the Euclidean region.
Additionally, \Caravel{} contains the \textbf{IntegralsBH} module, which provides a
large collection of one-loop integrals up to $O(\epsilon^0)$ (including massive
and massless propagators) which can be evaluated in up-to quad-double
precision using the \texttt{QD}~\cite{QD} package. This library has been adapted
from the \texttt{BHlibMassive} library employed in \cite{Anger:2017glm}.
\item\textbf{Coefficient Providers}: Two different modules perform the
hierarchical extraction of master-integrand coefficients via the
cut equations (see \cref{sec:cut_equations}).
The \textbf{AmpEng}
provider computes general one-loop integral coefficients, building up the
hierarchy of diagrams $\Delta$ in an automated fashion. For two-loop
calculations, the \textbf{AmplitudeCoefficients} module is employed. For a given
amplitude, it requires an input data file, which we call the
\textit{process library}. These process libraries contain all hierarchical
relations between the propagator structures in
the amplitude, as well as information about the color
decomposition~\cite{Ochirov:2016ewn,Ochirov:2019mtf} and the $D_s$-dependence based
on the particle content~\cite{Anger:2018ove,Abreu:2019odu,Sotnikov:2019onv}.
In this release, we provide the process libraries employed for the calculations in
\cite{Abreu:2017xsl,Abreu:2017hqn,Abreu:2018jgq,Abreu:2018zmy,Abreu:2019odu,
Abreu:2020lyk},
which were produced with private computer-algebra code.

  \item \textbf{Other modules}: further minor modules include miscellaneous
functionalities, for example some phase-space generators (including
momentum twistor parametrizations~\cite{Hodges:2009hk}) and information on the 
pole structure of relevant amplitudes.
\end{itemize}

All modules in \Caravel{} are implemented according to the concept of generic 
programming,
in which algorithms are designed to operate on any data type 
satisfying certain (minimal) requirements. In particular, our algorithms are 
well equipped to work with any numerical type, such as floating point number 
(of fixed or variable size) or numbers in an algebraic field (the rational
numbers or numbers in a finite field).
This allows us to 
perform evaluations of amplitudes with different fixed-size floating 
point numbers and the reconstruction of their analytic form from exact evaluations 
over finite fields with essentially a single implementation.

\section{Installation and setup}
\label{sec:install}

The source code of \Caravel{} can be obtained from a \texttt{git} repository at
\begin{center}
  \url{https://gitlab.com/caravel-public/caravel.git}.
\end{center}
\Caravel{} employs the \texttt{meson}~\cite{meson} build system,
which relies on \texttt{pkg-config} for resolving  dependencies. 
For more details on dependence resolution, configuration options and the
installation of optional libraries see the \texttt{README.md} and \texttt{INSTALL.md} files in the repository.
To build \Caravel{} in the default configuration and install it to the directory 
\verb|<install dir>| one can proceed as follows:
\begin{lstlisting}[numbers=left,caption=Obtaining and building \Caravel{} in 
its default configuration., label=lst:building, xleftmargin=\parindent]
> git clone https://gitlab.com/caravel-public/caravel.git
> cd caravel
> mkdir build 
> cd build 
> meson .. -D prefix=<install dir>
> ninja
> ninja install
\end{lstlisting}
All available test suites can be run with the command
\begin{lstlisting}
  > ninja test
\end{lstlisting}
executed in the build directory. Note that, depending on the hardware and build configuration, running 
all test suites can take a considerable amount of time. This is particularly noticeable for the first
time tests are run which produce and store warm-up information.

The default configuration of \Caravel{} provides very limited 
functionalities. This allows users to customize the installation to suit 
their particular needs, and additional configuration options and their current 
values can be queried by running 
\begin{lstlisting}
  > meson configure
\end{lstlisting}
in the build directory.
An option \verb|<option>| can be set to a particular value \verb|<value>| with 
the command
\begin{lstlisting}
  > meson configure -D <option>=<value>
\end{lstlisting}
These options can be set either at the configuration stage (step 5 of 
\cref{lst:building}), or any time before the building is initiated (step 6 of 
\cref{lst:building}). It is possible to specify several options at the same time.

Some of the options, listed first when running \texttt{meson configure}, 
are related to generic \Cpp{} compiler and 
linker options, which are automatically provided by the \texttt{meson} build 
system. These options are intended mostly for developers.

Options specific to \Caravel{} can be found in the section 
\emph{Project options}, which enable additional features.
We first describe the set of options most relevant for a user of the example
programs. 
\begin{itemize}
  \item \textbf{double-inverter}: Switch between 
  \texttt{Eigen} and \texttt{Lapack} for solving linear systems in double 
  precision.

  \item \textbf{finite-fields}: Enable 
  computation using finite fields. 
  Requires the external library \texttt{GMP}. This option can be set to 
  \texttt{false} (default) or \texttt{true}.

  \item \textbf{field-ext-fermions}: Enable the exact computation of master integral 
  coefficients for amplitudes involving fermions. Off-shell currents 
  are then evaluated in an algebraic extension of the number field
  in order to handle square roots originating from the solutions of quadratic 
  constraints for on-shell momenta (see \cref{sec:cut_equations}). The evaluation of cuts in the algebraic 
  extension is slower than in the corresponding field.
  For this reason, if only amplitudes in Yang-Mills theory are of interest, 
  the option should be left at the default value \texttt{false}. 

  \item \textbf{gravity-model}: Enable/select gravity models. 
  Possible choices are \texttt{none}, \texttt{Cubic}, \texttt{EH},
  \texttt{EH-GB-R3} and \texttt{all}. The last two options increase compilation times considerably. 
  These options give access to the computations of ref.~\cite{Abreu:2020lyk}, in particular to
  our implementation of the cubic formulation of the Einstein-Hilbert Feynman rules of~\cite{Cheung:2017kzx}.

  \item \textbf{precision-QD}: Enable the computation of master-integral
  coefficients and integrals in double-double (\texttt{-D precision-QD=HP}) or quad-double
  (\texttt{-D precision-QD=VHP}) floating-point precision. 
  In this case floating-point types are provided by the
  \texttt{QD} library~\cite{QD} and linear systems are solved with
  \texttt{Eigen}~\cite{Eigen}. Allowing computations in both double-double and quad-double precision 
  requires setting \texttt{-D precision-QD=all}. The default is
  \texttt{none}.

  \item \textbf{integrals}: Choose whether or not to compile the master
  integrals of the module \textbf{IntegralLibrary}. 
  The default is \texttt{none} which means that no master integrals are 
  compiled. If \texttt{goncharovs} is selected, 
  \texttt{GiNaC}~\cite{Vollinga:2005pk} is required. If 
  \texttt{pentagons} is selected the 
  \texttt{PentagonLibrary}~\cite{Gehrmann:2018yef} is required.\footnote{The modified version of the \texttt{PentagonLibrary} that is employed in \Caravel{} can be obtained 
from \url{https://gitlab.com/caravel-public/pentagon-library.git}. Notice the recent release of the
new \texttt{PentagonFunctions++} library~\cite{Chicherin:2020oor}.} The choice \texttt{all} 
  compiles both representations of the master integrals. 

  \item \textbf{lapack-path}: If necessary, specifies the path of
  \texttt{Lapack} if \texttt{meson} is unable to find the path to the library.
\end{itemize}

Beyond these, there are a number of options mainly useful for development:
\begin{itemize}
  \item \textbf{caravel-debug}: Enable a dynamic handling of debugging information from specific source files.
  To use this feature, simply place a file named \emph{debug.dat} in the directory
  where the corresponding binary is run. The file should contain 
  the filenames of the source files (without the full path to it), 
  which should provide additional debugging information. One filename should
  be listed per line and lines starting with \texttt{\#} are ignored. 
  This option can be set to 
  \texttt{false} (default) or \texttt{true}.

  \item \textbf{doxygen}: Enable the generation of HTML API
  documentation. This requires \texttt{Doxygen}~\cite{doxygen}.
  This option can be set to 
  \texttt{false} (default) or \texttt{true}. 

  \item \textbf{ds-strategy}: Select the algorithm for the reconstruction of
  the dependence of the two-loop amplitudes on the dimensional regulator $D_s$
  (see \cref{sec:integrands}).
  Possible values are \texttt{decomposition} (default), referring to the decomposition 
  by particle content, and \texttt{sampling}, referring to the reconstruction 
  of the $D_s$ polynomial coefficients from the sample values. 
  The former provides a significant efficiency boost 
  so we recommend not to change its default value. The option
  \texttt{decomposition} is currently not supported for gravity amplitudes.

  \item \textbf{instantiate-rational}: Enable selected computations with rational 
  numbers. This option requires \textbf{finite-fields} set to \texttt{true}. 
  This option can be set to \texttt{false} (default) or \texttt{true}.

  \item \textbf{precision-arbitrary}: Enable computation with arbitrary-precision floating-point types
  included in the \texttt{GMP} and \texttt{MPFR} libraries. This option can be set to 
  \texttt{false} (default) or \texttt{true}.

  \item \textbf{timing}: Enable the printout of the
  time spent in different contributions to amplitude's calculations 
  at the end of each program. This option can be set to \texttt{false} (default) or \texttt{true}.
\end{itemize}
Enabling some of these features introduces dependencies on third-party 
libraries, which the user should make available before the start of the building
process. Since certain options may significantly increase build times, 
we suggest to enable only the features necessary for each calculation. 
Depending on the chosen configuration, building 
\Caravel{} can take from a few minutes up to half an hour on a modern 
multi-core processor.

\section{Example programs}
\label{sec:run}
In this section we introduce the example programs provided with this release,
which demonstrate the main features of \Caravel{}. 
For each one we specify the required configuration setup, a brief explanation 
of the computations it performs and instructions for execution.
More information about these programs can be found in the file
\verb|examples/README.md|, contained in the \Caravel{} repository.
All programs can be found in \verb|build/examples|, where
\verb|build| is the build directory created in \cref{lst:building}.
Before turning to the description of each example, we first establish some
conventions and explain the general structure of the command-line input 
that the user must provide.

\subsection{Helicity amplitudes}

Each particle $q$ in a multi-particle helicity amplitude is labelled
by the particle type $f_q$ (here, $f_q$ can be a gluon, an (anti)quark, or a graviton), 
the helicity state $h_q$ and the color index $c_q$ (for color-charged particles).
An $n$-particle amplitude depends on all this data, that is
\begin{equation}\label{eq:ampDef}
    \mathcal{M}_n\equiv\mathcal{M}_n(1_{f_1}^{c_1, h_1}, \ldots, {n}_{f_{n}}^{c_{n}, h_{n}}).
\end{equation}
We assign a momentum index $q$ to particle $q$.
All particles and their momenta are considered outgoing.
Note that the color indices $c_q$ are not present in pure graviton amplitudes.
We consider the perturbative expansion of the (bare) helicity amplitudes and write
\begin{equation}\label{eq:pertExp}
  \mathcal{M}_n = a_0^{\lambda} \left( \mathcal{M}_n^{(0)} + \left(\frac{a_0}{4\pi}\right)^2 \mathcal{M}_n^{(1)} + \left(\frac{a_0}{4\pi}\right)^4 \mathcal{M}_n^{(2)} + \cdots \right).
\end{equation}
where $a_0$ is a generic bare coupling constant and $\lambda$ denotes
the power of the leading-order amplitude. For QCD amplitudes, the expansion
is in powers of the strong coupling, i.e.~$a_0 = g_0$, where $g_0 = \sqrt{4\pi \alpha_s}$.
For Einstein gravity amplitudes, instead, $a_0 = \kappa_0/2$, 
where $\kappa_0 = \sqrt{32 \pi G_N}$ and $G_N$ is Newton's constant.
In this section we will often refer to the index $L$ in $\mathcal{M}_n^{(L)}$
as the loop order, as for the examples we will consider the two numbers
are aligned. $L=0$ corresponds to  tree-level amplitudes.

As already stated in section \ref{sec:methods}, see in 
particular eq.~\eqref{eq:colorDec}, \Caravel{} computes
the coefficients $\mathcal{A}_n^{(L)}$ of
the decomposition of $\mathcal{M}_n^{(L)}$ in terms of a set of color structures. 
For  gravity amplitudes, this decomposition is trivial and
\Caravel{} directly computes the $\mathcal{M}_n^{(L)}$.
More generally, to properly define the helicity amplitudes $\mathcal{A}_n^{(L)}$
we must specify several of our conventions and this is done in detail
in \ref{sec:helicity_caravel}. These conventions are important
for defining the output of the example programs described in this section.

The example programs we provide compute tree-level, one-loop and two-loop 
amplitudes. While the tree-level example program allows one to evaluate amplitudes
with an arbitrary number of particles, the one-loop and two-loop
example programs evaluate at most five-point amplitudes, 
up to order $\epsilon^2$ for one loop and $\epsilon^0$ for two loops.
The integral normalization for the amplitudes computed numerically in the
example programs is defined by equation~\eqref{eq:pertExp}. We stress that
this differs from the internal normalization in eq.~\eqref{eq:normInt}.
Indeed, the integrals in the example programs are normalized as
\begin{equation}
 \mathcal{I}_{\Gamma,i} = \int \left( \prod_{j=1}^L\frac{d^D\ell_j}{(2\pi)^D}\right)\,
\frac{m_{\Gamma,i}(\ell_l)}{\prod_{k\in P_\Gamma} \rho_k(\ell_l)}.
  \label{eq:normExample}
\end{equation}
Additionally, the QCD loop amplitudes are evaluated in the leading-color
limit of QCD.
In this limit, we keep the leading term of~eq.~\eqref{eq:colorDec} in the limit of a large number
of colors $N_c$, but consider the ratio $N_f/N_c$ to be fixed, where $N_f$ is the number
of massless flavors. The leading-color amplitudes have a decomposition
in terms of powers of this ratio, specifically
\begin{equation}
\mathcal{A}^{(L)}(1_{p_1}^{h_1},\ldots,n_{p_n}^{h_n}) = \mathcal{A}^{(L)[0]} + \frac{N_f}{N_c}\mathcal{A}^{(L)[1]} + \ldots + \left(\frac{N_f}{N_c}\right)^L\mathcal{A}^{(L)[L]}.
\label{eqn:partialamp}
\end{equation}

\subsection{Specifying program input}\label{sec:input}
Many of the programs that we provide can be run for a variety of different
scattering amplitudes. 
To evaluate different amplitudes,
we provide a uniform interface by passing (arbitrarily-ordered)
command-line arguments.
These arguments are formatted similarly to \texttt{Mathematica} lists with heads.

\paragraph{Specifying particles}
In the command-line interface, a particle is specified as
\begin{lstlisting}
Particle[field, index, state]
\end{lstlisting}
A (momentum) \texttt{index} needs to be provided for each particle,
starting from $1$ up to the number of particles in the scattering
process.
Gluons are specified by setting the \texttt{field} to \texttt{gluon}, and  
\texttt{state} should be set to either \texttt{p} ($+$ helicity) or \texttt{m} ($-$ helicity).
For massless quarks, \Caravel{} offers multiple possibilities. Generic unflavored  
quarks and their anti-particles are input as \texttt{q} and \texttt{qb} 
respectively. Fields of definite flavor can also be specified with \texttt{u,d,c,s,b} and 
\texttt{ub,db,cb,sb,bb}. We note that  scattering processes 
with identical fermion flavors are currently not supported. However, these can be obtained
from anti-symmetrizing distinct flavor amplitudes 
(see for example~\cite{Bern:1996ka,Glover:2004si,DeFreitas:2004kmi}).
The associated \texttt{state}s for fermions are labeled with \texttt{qbp} 
(+ helicity) and \texttt{qm} (-- helicity). The graviton
field is labelled by \texttt{G} and its polarization states are given by
\texttt{hmm} (--~-- helicity) and \texttt{hpp} (++ helicity). In table
\ref{tab:ParticleStatePairings} we summarize the available \texttt{field} and 
\texttt{state} options to define particles in \Caravel{}.
\begin{table}[htb]
\begin{center}
\begin{tabular}{c | c | c}
  Type & \texttt{field} & \texttt{state} \\
  \hline
  Gluon     & \texttt{gluon} & \texttt{p}, \texttt{m} \\
  Quark     & \texttt{q}, \texttt{u}, \texttt{d}, \texttt{c}, \texttt{s}, \texttt{b} & \texttt{qbp}, \texttt{qm} \\
  Anti-quark & \texttt{qb}, \texttt{ub}, \texttt{db}, \texttt{cb}, \texttt{sb}, \texttt{bb} & \texttt{qbp}, \texttt{qm} \\
  Graviton  & \texttt{G} & \texttt{hpp}, \texttt{hmm} 
\end{tabular}
\end{center}
\caption{Allowed \texttt{field}/\texttt{state} pairings in a \texttt{Particle} list.}
\label{tab:ParticleStatePairings}
\end{table}

\paragraph{Specifying amplitudes}
The simplest scattering amplitudes are color-ordered tree-level helicity 
amplitudes $\mathcal{A}^{(0)}(1_{f_1}^{h_1}, \ldots, n_{f_n}^{h_n})$. Such a 
tree-level amplitude is specified in \Caravel{} by an ordered list of particles
\begin{lstlisting}
Particles[Particle[..],Particle[..],Particle[..],..]
\end{lstlisting}
where each particle is defined as in the previous section. Note that 
color-ordered amplitudes are invariant under cyclic permutations of the external
particles. For gravity the same interface is used, however the ordering of the
external gravitons does not matter.

For loop amplitudes, 
the partial amplitude $\mathcal{A}^{(L)[k]}$ of equation \eqref{eqn:partialamp} can
be specified in the command-line interface by 
\begin{lstlisting}
PartialAmplitudeInput[Particles[Particle[..], ..], NfPower[k]]
\end{lstlisting}
where \texttt{NfPower[k]} corresponds to the desired power of $N_f$.
This entry is optional, and if it is omitted then the $N_f^0$ contribution
is computed. For gravity amplitudes this entry is meaningless and 
should be omitted. In table~\ref{tab:PartialAmplitudeInput}
we give examples of valid  $\texttt{PartialAmplitudeInput}$.

\begin{table}[ht]
  \begin{center}
\footnotesize
\begin{tabular}{c|c}
Amplitude & \texttt{PartialAmplitudeInput} \\
\hline
$\mathcal{A}^{(2)[2]}(1_g^+, 2_g^+, 3_g^+, 4_g^+, 5_g^+)$ & 
\begin{lstlisting}
"PartialAmplitudeInput[
  Particles[
    Particle[gluon,1,p], 
    Particle[gluon,2,p], 
    Particle[gluon,3,p], 
    Particle[gluon,4,p], 
    Particle[gluon,5,p]
  ],
  NfPower[2]
]"
\end{lstlisting} \\[1em]
  \hline
$\mathcal{A}^{(L)[0]}(1_q^+, 2_{\overline{q}}^-, 3_g^+, 4_g^-)$ &
\begin{lstlisting}
"PartialAmplitudeInput[
  Particles[
    Particle[q,1,qbp], 
    Particle[qb,2,qm],
    Particle[gluon,3,p], 
    Particle[gluon,4,m]
  ]
]"
\end{lstlisting} \\
  \hline
$\mathcal{A}^{(L)[0]}(1_u^+, 2_{\overline{u}}^-, 3_{d}^+, 4_{\overline{d}}^-)$  &
\begin{lstlisting}
"PartialAmplitudeInput[
  Particles[
    Particle[u,1,qbp], 
    Particle[ub,2,qm],
    Particle[d,3,qbp], 
    Particle[db,4,qm]
  ]
]"
\end{lstlisting}

\end{tabular}
\end{center}
\label{tab:PartialAmplitudeInput}
\caption{Examples of valid \texttt{PartialAmplitudeInput} lists, all related to leading-color QCD amplitudes. The quotation marks are required to keep the line breaks in a shell execution.}
\end{table}
%

\paragraph{Specifying kinematics}

By default, most of the examples we provide evaluate 
amplitudes on phase-space points which allow one to 
reproduce the results of ref.~\cite{Abreu:2018jgq}. 
Nevertheless, the user can request evaluations at different
phase-space points. Since internally most of the example programs
perform calculations in a finite field, one has to be sure that the momenta associated with the 
chosen phase-space point can be represented in a finite field. Finding
such points is in general a non-trivial problem. For all the examples
we will be concerned with, however, it can be solved by using
a special parametrization of phase-space, called \emph{momentum twistor} parametrization~\cite{Hodges:2009hk}.
In \ref{app:twistor}, we give more details on this parametrization for four- and five-point
massless kinematics. In particular, we give equations that relate our choice of twistor
parameters to Mandelstam variables $s_{ij}=(p_i+p_j)^2$, where
$p_i$ denotes the momentum of the external particle with particle index $i$.

To evaluate amplitudes at a chosen phase-space point, the user should
provide a list with the head \texttt{TwistorParameters}.
For four-point kinematics, the phase space is directly parametrized by the Mandelstam variables 
\texttt{s12} and \texttt{s23} 
which are passed as
\begin{lstlisting}
TwistorParameters[s12, s23]
\end{lstlisting}
For five-point kinematics the parametrization is given in terms of the 5 independent twistor
parameters \texttt{x0}, $\ldots$, \texttt{x4}, see \ref{app:twistor} for more details. 
These are passed as
\begin{lstlisting}
TwistorParameters[x0, x1, x2, x3, x4]
\end{lstlisting}

\subsection{Numerical amplitude evaluation}
In this subsection we present a series of programs which allow the numerical
evaluation of a number of QCD scattering amplitudes
at tree level, one loop and two loop as well as graviton amplitudes at tree level.

\subsubsection{Tree level}
The program \texttt{treeamp} evaluates tree-level amplitudes for a variety of
processes. It can be executed by specifying the corresponding amplitude with a \texttt{Particles} list~(see section~\ref{sec:input}). For example:
\begin{lstlisting}
 > ./treeamp "Particles[Particle[..],...,Particle[..]]"
\end{lstlisting}
The program randomly generates a phase-space point and prints the value of the
specified tree-level amplitude as well as the point. In \ref{app:wavefunctions}
we provide details about the normalization employed for external helicity states,
which are necessary to specify our phase conventions.

If \texttt{treeamp} is to be used to evaluate $N$-graviton scattering amplitudes, \Caravel{} must be configured with the option
\begin{lstlisting}
-D gravity-model=Cubic
\end{lstlisting}

Instead of evaluating the amplitudes on randomly-generated
phase-space points the user can also provide external momenta by placing a file
with the name \texttt{treeampPSP.dat} into the same directory as the executable. The file should
list one momentum per line in the format: 
\begin{lstlisting}
E PX PY PZ
\end{lstlisting}
\texttt{treeamp} will find this file, read the momenta and, after performing on-shell and 
momentum-conservation checks, will compute the corresponding amplitude.

If \Caravel{} has been configured to include high-precision floating-point types (with the 
option \texttt{precision-QD} set to \texttt{HP}, \texttt{VHP} or \texttt{all}),
then high-precision amplitudes can be computed by passing an additional
command-line input
\begin{lstlisting}
 > ./treeamp "Particles[...]" "HighPrecision[prec]"
\end{lstlisting}
where \texttt{prec} is either \texttt{HP} for double-double or \texttt{VHP} for
quad-double precision. 

Finally, the program can be used to compute tree amplitudes using finite-field arithmetic (\texttt{-D finite-fields=true} required).
In this case the cardinality of the finite field has to be passed as a parameter to the program
\begin{lstlisting}
> ./treeamp "Particles[...]" "Cardinality[p]"
\end{lstlisting}
where \texttt{p} is a prime number smaller than $2^{31}$ and larger than $2^{30}$, such
as $\texttt{p}=2^{31}-1=2147483647$. For finite-field 
evaluations, the program does not randomly generate phase-space points and so
the user must provide a set of valid external momenta using a file
named \texttt{treeampPSP.dat} as explained above. Notice that in the case of 
finite-field evaluations we use a space-time metric with alternating 
signature $(+,-,+,-)$ to enhance performance.
Thus, the momentum components should be provided in this signature.
For example, using \texttt{treeampPSP.dat} we can pass the rational 
phase-space point
\begin{verbatim}
  1/3        1/3   -2        2
 -5/16       1/4   -9/64    15/64
329/144  -355/144  17/16   -25/48
-83/36    271/144  69/64  -329/192
\end{verbatim}
to evaluate four-point amplitudes for the momenta
\begin{align*}
 &p_1 = \left( \frac{1}{3}, \frac{1}{3}, -2, 2\right)\;, \qquad 
 &&p_2 = \left( -\frac{5}{16}, \frac{1}{4}, -\frac{9}{64}, \frac{15}{64}\right)\;, \\
 &p_3 = \left( \frac{329}{144}, -\frac{355}{144}, \frac{17}{16}, -\frac{25}{48}\right)\;, \qquad 
 &&p_4 = \left( -\frac{83}{36}, \frac{271}{144}, \frac{69}{64}, -\frac{329}{192}\right)\;, 
\end{align*}
corresponding to $s_{12}=-3/4$ and $s_{23}=-1/4$. The program 
converts the rational momentum components into their image in the chosen finite field.

\subsubsection{One-loop amplitude \texorpdfstring{to $\mathcal{O}(\epsilon^2)$}{}}
\label{subsec:oneloopEval}
The program \texttt{amplitude\_evaluator\_1l} numerically evaluates one-loop four- and 
five-parton helicity amplitudes up to
order $\mathcal{O}(\epsilon^2)$ in the dimensional regulator.
The master-integral coefficients 
are rationally reconstructed from finite-field evaluations. 
The integrals are then computed (in double precision) to obtain the 
numerical value for the amplitude. 
If the corresponding tree-level amplitude is non-vanishing the one-loop result
is normalized by this tree-level amplitude. Otherwise,
the result is normalized by a spinor weight as defined in \ref{sec:spinor_weight},
see in particular eq.~\eqref{eq:swNorma}.
The Laurent expansion of the amplitude is printed to the standard output.

By default, the program runs on the phase-space points defined in 
ref.~\cite{Abreu:2018jgq} so that the user can easily reproduce the results
in tables $3$ and $4$. In the current implementation, this example program runs only 
in the Euclidean region of phase space as \Caravel{} does not include the analytic
continuation of the integrals. 
We note that the evaluation of five-point amplitudes
takes a considerable amount of time because of the evaluation of
the one-loop pentagon integral through order $\epsilon^2$.

In order to enable this example, \Caravel{} has to be configured with the 
following options:
\begin{lstlisting}
-D finite-fields=true
-D field-ext-fermions=true
-D integrals=all
\end{lstlisting}
The program can be executed by passing the appropriate amplitude input and
kinematic point specifications. For example:
\begin{lstlisting}
> ./amplitude_evaluator_1l "PartialAmplitudeInput[Particles[ 
 Particle[gluon,1,m], Particle[gluon,2,m],  
 Particle[gluon,3,p], Particle[gluon,4,p]]]" \
 "TwistorParameters[-1/3, -1/5]"
\end{lstlisting}
evaluates the one-loop color-ordered helicity amplitude 
$\mathcal{A}^{(1)[0]}(1_g^-,2_g^-,3_g^+,4_g^+)$ at  $s_{12}=-\frac{1}{3}$ 
and $s_{23}=-\frac{1}{5}$.

\subsubsection{Two-loop amplitude}
The program \texttt{amplitude\_evaluator\_2l} numerically evaluates two-loop four-
and five-parton helicity amplitudes to $\mathcal{O}(\epsilon^0)$ in the
dimensional regulator. It works in the same way as the one-loop
program described above. The master-integral coefficients are
rationally reconstructed from finite-field evaluations and subsequently 
combined with the master integrals into a semi-analytic object, which is 
expressed in terms of unevaluated special functions.
Next, those special functions are evaluated to produce the amplitude
whose Laurent expansion is then printed to the standard output.
The computation of the master-integral coefficients is performed in parallel,
using all available threads in the CPU, then the special functions are evaluated
sequentially in a single thread. As at one-loop, the amplitude is normalized
either by the corresponding non-vanishing tree-level amplitude or by the
spinor-weight defined in \ref{sec:spinor_weight},
see in particular eq.~\eqref{eq:swNorma}. On the first evaluation of
each amplitude, the program performs a warm-up run, as described in sections
\ref{sec:cut_equations} and \cref{sec:integrands}.

By default, the program runs on phase-space points defined in 
ref.~\cite{Abreu:2018jgq} and can be used to reproduce the results presented 
in the tables $1$ and $2$. 
As at one-loop, this example program is restricted to the
Euclidean region of phase space, as master integrals are so far included only
for this region.

In order to enable this example the \Caravel{} library has to be configured with
the following options:
\begin{lstlisting}
-D finite-fields=true
-D field-ext-fermions=true
-D integrals=all
\end{lstlisting}
The program can be executed in an analogous way to what was described in the
previous example in \cref{subsec:oneloopEval}. Additionally, the user can pass the
argument \texttt{Verbosity[All]} in order to print to the standard
output extra information on the computations performed. For example:
\begin{lstlisting}
> ./amplitude_evaluator_2l "PartialAmplitudeInput[Particles[ 
 Particle[gluon,1,m], Particle[gluon,2,m], 
 Particle[gluon,3,p], Particle[gluon,4,p]],NfPower[1]]" \
 "TwistorParameters[-1/3, -1/5]"
\end{lstlisting}
evaluates the two-loop color-ordered helicity amplitude 
$\mathcal{A}^{(2)[1]}(1_g^-,2_g^-,3_g^+,4_g^+)$ for $s_{12}=-\frac{1}{3}$ and $s_{23}=-\frac{1}{5}$.

The evaluation of two-loop amplitudes is considerably more involved than the
corresponding one-loop amplitudes. For example, the runtime of the
most complex two-loop five-parton amplitude is of the order of 12 minutes to rationally
reconstruct all master-integral coefficients on the default phase-space point (using 22 finite-field
evaluations), while the computation of the pentagon functions in double
precision by the external library provided with ref.~\cite{Gehrmann:2018yef} takes about 23 minutes (employing a modern 12-core Intel i7 processor).

\subsubsection{Five-point two-loop finite remainder}

The program \texttt{finite\_remainder\_2l} numerically computes the finite remainder of
planar two-loop five-parton amplitudes. 
The numerical calculation of finite remainders was instrumental in
order to reconstruct the analytic form of the planar two-loop five-parton 
amplitudes in refs.~\cite{Abreu:2018zmy,Abreu:2019odu}.
The program proceeds by building the requested two-loop amplitude and its
corresponding infrared subtraction, which also requires the one-loop 
amplitude to $\mathcal{O}(\epsilon^2)$. The finite remainder obtained in this way
is decomposed in terms of the special functions introduced in ref.~\cite{Gehrmann:2018yef}.
As in the two-loop amplitude evaluation example,
the computation of the two-loop master integral coefficients is
performed in parallel, using all available threads in the CPU
and, on the first evaluation, a warm-up run will be performed for the two-loop
amplitude. The program automatically verifies that all poles in $\epsilon$
cancel exactly, rationally reconstructs the special function coefficients in the
finite remainder and subtraction term, and evaluates the special functions. The
program then prints the values of the subtraction term, the remainder and the amplitude.
We refer to ref.~\cite{Abreu:2019odu} for the precise definition of our
subtraction conventions. We note that the run-time of this program is
comparable to that of the two-loop amplitude evaluation program.

In order to enable this example \Caravel{} has to be configured with
the following options:
\begin{lstlisting}
-D finite-fields=true
-D field-ext-fermions=true
-D integrals=pentagons
\end{lstlisting}
This example can also be enabled with the \texttt{integrals} option set to \texttt{-D integrals=all}.
The program has similar command-line arguments as in the examples above.
For example, it can be executed with 
\begin{lstlisting}
> ./finite_remainder_2l "PartialAmplitudeInput[Particles[ 
 Particle[gluon,1,p],Particle[gluon,2,p],Particle[gluon,3,p], 
 Particle[gluon,4,p],Particle[gluon,5,p]],NfPower[2]]"
\end{lstlisting}
in order to compute the finite remainder for the color-ordered two-loop 
amplitude $\mathcal{A}^{(2)[2]}(1_g^+,2_g^+,3_g^+,4_g^+,5_g^+)$. As in the two previous
examples, the phase-space point can be specified in the command-line input as a list with the 
head \texttt{TwistorParameters}.
The optional argument \verb|Verbosity[All]| or \verb|Verbosity[Remainder]| can 
be passed in the command line to request the printing of additional information.

\subsection{Analytic reconstruction of amplitudes}\label{sec:analyticReconstruction}
An important application of the numerical techniques for amplitude calculation
that \Caravel{} provides is the reconstruction of analytic results from numerical
evaluations. Here we present two programs which reconstruct either two-loop 
four-parton or one-loop five-parton amplitudes from numerical evaluations.
These examples rely on \texttt{MPI} for parallelization.

\subsubsection{Program output}
\label{sec:analyticOutput}
The two analytic reconstruction example programs share common output features,
which we describe here. Both programs perform the analytic reconstruction over a single
finite field and attempt to rationally reconstruct the result. The computation
in the finite field is saved as a text file in the local directory
\texttt{analytics/amplitudes\_XY/}, where \texttt{XY} refers to the relevant
cardinality. The master-integral coefficients are then
rationally reconstructed employing only the single finite-field evaluation performed.
This rational reconstruction is  cross checked against a numerical computation at a single phase-space point
in a second finite field. In the case of a successful reconstruction, the amplitude is
saved in a text file under \texttt{analytics/amplitudes\_rational/}.
The reconstructed amplitudes are normalized
either by the corresponding non-vanishing tree-level amplitude or by the spinor weight
defined in \ref{sec:spinor_weight},
see in particular eq.~\eqref{eq:swNorma}.

The files produced by each program are named according to the requested
\texttt{PartialAmplitudeInput}. They contain a string that encodes
the decomposition of the amplitude as a linear combination of
master integral coefficients and master integrals.\footnote{The format
used in this string allows it to be directly imported into
Mathematica for usage if desired.}
Each master integral is specified by a string of the form
\begin{verbatim}
Topology[NumeratorLabel, {D1, D2, .., Dm}].
\end{verbatim}
Here, \texttt{Topology} is a human readable name for the topology, e.g.~\texttt{Triangle} 
or \texttt{DoubleBox}. The list of \texttt{Di} is the list
of inverse propagators of the master integral, written in terms of
the loop momenta (\texttt{l1}, \texttt{l2}) and external momenta (\texttt{k1},
\texttt{k2}, $\ldots$) of the amplitude. Our conventions in \Caravel{} are that
all external momenta are outgoing.
The label \texttt{NumeratorLabel} denotes the numerator of the master integral.
For the example programs provided with this release, there are four possible
values for this label, which we list in table~\ref{tab:MasterNumerators}.
For one-loop integrals, these include numerators built from powers
of $\mu^2$, which is the scalar product of the $(D-4)$ dimensional components of the loop
momentum, i.e.
\begin{equation}
    \mu^2 = \ell^{(D-4)} \cdot \ell^{(D-4)}.
\end{equation}

\begin{table}[]
  \begin{center}
\begin{tabular}{c|c}
\texttt{NumeratorLabel} & Master-integral numerator \\
\hline
\texttt{scalar} & $1$ \\
\texttt{mu2} & $\mu^2$ \\
\texttt{mu4} & $(\mu^2)^2$ \\
\texttt{dbTensor} & $(p_1+\ell_2)^2$ \\
\end{tabular}
\end{center}
\caption{Table of numerator labels used in output files and their explicit
  expression in momentum space. The conventions for \texttt{dbTensor}
  are specified in fig.~\ref{fig:dbox}.}
  \label{tab:MasterNumerators}
\end{table}

\begin{figure}
\centering
  \includegraphics[scale=.5]{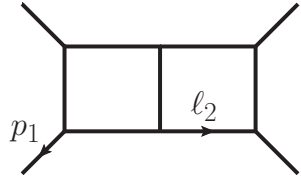}
  \caption{Specifying momentum routing for the double-box tensor integral.}
  \label{fig:dbox}
\end{figure}

In order to aid reading of the output, we provide a collection of Mathematica
routines in \texttt{math/CaravelGraph.m} which produce graphical repesentations
of the integrals. The output format as described above is not appropriate for
these routines. The example programs therefore also writes a textfile in the
directory \texttt{analytics/integral\_info/}, which contains a list of
replacement rules that allow one to use these routines.

\subsubsection{Univariate amplitude reconstruction}
As a simple example of the two-loop analytic reconstruction capabilities of
\Caravel{}, the program \texttt{4parton\_2loop\_analytics\_MPI} analytically 
computes the reduction to master integrals. The massless four-point amplitudes 
depend only on the Mandelstam variables $s=(p_1+p_2)^2$ and $t=(p_2+p_3)^2$. By
setting $s=1$ and $x=t/s$ the amplitude depends only on a single parameter. Its
analytic dependence on $x$ is  reconstructed from exact numerical evaluations
of the master-integral coefficients over a finite field, which are then fed 
into Thiele's interpolation formula. The dependence on $s$ is then
recovered by dimensional analysis.

To enable this example, \Caravel{} has to be configured with the options
\begin{lstlisting}
-D finite-fields=true
-D field-ext-fermions=true
\end{lstlisting}
To compute the two-loop four-gluon all-plus-helicity amplitude,
for example, the program should be executed with
\begin{lstlisting}
> mpirun -np <ncores> ./4parton_2loop_analytics_MPI \
  "PartialAmplitudeInput[Particles[Particle[gluon,1,p], 
  Particle[gluon,2,p],Particle[gluon,3,p], 
  Particle[gluon,4,p]]]"
\end{lstlisting}

\subsubsection{Multivariate amplitude reconstruction}
The program computes the analytic form of five-parton one-loop amplitudes,
using multivariate functional reconstruction. The five-parton amplitudes depend
on five twistor parameters $x_0,\ldots,x_4$, see \ref{app:twistor}. 
The problem is reduced to a four-dimensional
reconstruction by setting $x_4 = 1$. The dependence on $x_4$ is recovered
from dimensional analysis.

In order to enable this example, \Caravel{} has to be configured with the options
\begin{lstlisting}
-D finite-fields=true
-D field-ext-fermions=true
\end{lstlisting}
To compute the one-loop five gluon all-plus-helicity amplitude,
for example,  the program should be executed with
\begin{lstlisting}
> mpirun -np <ncores> ./5parton_1loop_analytics_MPI \
  "PartialAmplitudeInput[Particles[Particle[gluon,1,p], 
  Particle[gluon,2,p],Particle[gluon,3,p],Particle[gluon,4,p], 
  Particle[gluon,5,p]]]"
\end{lstlisting}
%

\section{Conclusions}
\label{sec:conclusions}

We have presented \Caravel{}, a \Cpp{} framework for the computation of
multi-loop amplitudes through the multi-loop numerical unitarity method. This is the first
publicly available program of its kind. We have provided a series of example
programs which showcase the main functionalities of \Caravel{}. In particular,
these examples give access to all details of the calculation of the planar 
two-loop five-parton scattering amplitudes~\cite{Abreu:2018zmy,Abreu:2019odu} 
and the two-loop four-graviton scattering amplitude in 
Einstein gravity~\cite{Abreu:2020lyk}.

In its current form, \Caravel{} is not meant to be able to automatically
evaluate arbitrary two-loop multi-leg amplitudes. Rather, it is meant
as a framework to do so, which should be complemented
with process-specific information such as, for instance, the master integrals.
The modular fashion in which \Caravel{} is constructed allows one to easily
add these new features.
We aim to continue the development of \Caravel{}, to
increase the pool of multi-loop amplitude computations that it can perform, while 
also extending its autonomy.

\section*{Acknowledgments}

We gratefully acknowledge contributions to \Caravel{} made by 
Matthieu Jaquier during the initial stages of the development.
We also wish to thank Mao Zeng for discussions.
The work of S.A.~is supported by the Fonds de la Recherche Scientifique--FNRS, 
Belgium.
The work of F.F.C.\ is supported by the 
U.S. Department of Energy under grant DE-SC0010102.
The work of V.S.\ is supported by the European Research Council (ERC) 
under the European Union's Horizon 2020 research and innovation programme,
\textit{Novel structures in scattering amplitudes} (grant agreement No.\ 725110).
The work of B.P.~is supported by the French Agence Nationale
pour la Recherche, under grant ANR--17--CE31--0001--01.
M.S.R.'s work is funded by the German Research Foundation (DFG) within the 
Research Training Group GRK 2044.
The authors acknowledge support by the state of
Baden-W\"urttemberg through bwHPC.

\appendix

\section{Helicity amplitudes in Caravel}
\label{sec:helicity_caravel}

In this appendix we summarize several of our conventions, that allow us
to precisely define the helicity amplitudes computed by \Caravel{}.
In \ref{sec:colorDec} we discuss the conventions regarding the color
decomposition. In \ref{app:wavefunctions} we present our conventions for 
external helicity states. Finally, in \ref{sec:spinor_weight} we discuss the 
spinor-weight normalization.

\subsection{Color decomposition}
\label{sec:colorDec}

As already stated in section \ref{sec:methods}, see in 
particular eq.~\eqref{eq:colorDec}, \Caravel{} computes
the coefficients of $\mathcal{M}_n^{(L)}$ in a decomposition in terms of
color structures. 
For Einstein gravity amplitudes, this decomposition is trivial and
\Caravel{} directly computes the $\mathcal{M}_n^{(L)}$.
For QCD amplitudes, however, to properly define the objects computed in the example programs
we must specify the color decomposition and our conventions for
the color algebra. We follow the conventions of ref.~\cite{Dixon:1996wi}
and denote the
fundamental generators of the $SU(N_c)$ group by $(T^a)^{\;\bar{\jmath}}_{i}$,
where the adjoint index $a$ runs over $N_c^2-1$ values and the (anti-)
fundamental indices $i$ and $\bar \imath$ run over $N_c$ values. We use the
normalization $ \mathrm{Tr}(T^a T^b) = \delta^{ab}$. Finally, we define
$F^{abc}=\mathrm{Tr}\big([T^a, T^b,] T^c\big)$, which is closely
related to the $SU(N_c)$ structure constants.

\paragraph{Tree-level QCD amplitudes}
In order to define in a unified way the color-ordered tree-level amplitudes 
that we compute, we shall
consider QCD with all fields (that is, both quarks and gluons) in the adjoint representation. 
In this `adjoint QCD' theory,
amplitudes with any number of partons can be expressed in terms of $(n-2)!$
color-ordered partial amplitudes using the decomposition of ref.~\cite{DelDuca:1999rs},
\begin{align}\begin{split}
  \mathcal{M}^{(0)} = \sum_{\sigma\in S_{n-2}} &C(a_{n-1}, a_{\sigma_{1}}, 
  \ldots, a_{\sigma_{n-2}}, a_n) \\
  &\times
  \mathcal{A}^{(0)}({n\!-\!1}_{p_{n-1}}^{h_{n-1}}, {\sigma_{1}}_{p_{\sigma_{1}}}^{h_{\sigma_{1}}}, 
  \ldots, {\sigma_{n-2}}_{p_{\sigma_{n-2}}}^{h_{\sigma_{n-2}}}, {n}_{p_{n}}^{h_{n}}).
\label{eq:DDMDecomposition}
\end{split}\end{align}
Here, $S_n$ denotes all permutations of $n$ indices and $C$ is a color structure given by
\begin{equation}
C(a_1, \ldots, a_n) = F^{a_1 a_2 x_1} F^{x_1 a_3 x_2} \cdots 
F^{x_{n-4} a_{n-2} x_{n-3}} F^{x_{n-3} a_{n-1} a_n},
\end{equation}
with the $F^{abc}$ defined above. We stress that
we simply use adjoint QCD to define quark and gluon amplitudes in a unified
way. It is a well understood procedure to assemble the
multi-parton QCD amplitude from these color-ordered amplitudes, see for
example~\cite{DelDuca:1999rs, Johansson:2015oia, Ochirov:2019mtf}.

\paragraph{Leading-color QCD loop amplitudes}
Beyond tree-level, the example programs compute the color ordered amplitudes relevant for the
leading-color limit of QCD. In this limit, we keep the leading term for a large number
of colors $N_c$, but consider the ratio $N_f/N_c$ to be fixed, where $N_f$ is the number
of massless flavors.

We first discuss the four-point amplitudes and consider amplitudes for the
scattering of four gluons, one quark pair and two gluons, and two distinct quark
pairs. In the leading-color approximation we write
\begin{align}
  \begin{split}
  \label{eq:bareAmp4p}
  \mathcal{M}^{(L)}(1_g, 2_g, 3_g, 4_g) \big\vert_{\textrm{leading color}} = &
  N_c^L \sum_{\sigma\in S_4/Z_4} \mathrm{Tr}\left(
  T^{a_{\sigma(1)}} T^{a_{\sigma(2)}} 
  T^{a_{\sigma(3)}} T^{a_{\sigma(4)}} \right)\\
  & \times \mathcal{A}^{(L)}({\sigma(1)}_g, {\sigma(2)}_g, {\sigma(3)}_g, {\sigma(4)}_g)\,, 
  \end{split} \\
  \begin{split}
  \label{eq:ColorDec2Q4p}
  \mathcal{M}^{(L)}(1_q, 2_{\bar{q}}, 3_g, 4_g) \big\vert_{\textrm{leading color}} 
    = & N_c^L \sum_{\sigma\in S_2} 
  \left( T^{a_{\sigma(3)}} T^{a_{\sigma(4)}} \right)^{\;\bar{\imath}_2}_{i_1} \\
  & \times \mathcal{A}^{(L)}(1_q,2_{\bar{q}},\sigma(3)_g,\sigma(4)_g)\,,
  \end{split}\\
  \begin{split}
  \label{eq:ColorDec4Q4p}
  \mathcal{M}^{(L)}(1_q, 2_{\bar{q}}, 3_Q, 4_{\bar{Q}})
  \big\vert_{\textrm{leading color}} 
  = & 
  N_c^L \,\delta^{\;\bar{\imath}_{2}}_{i_{3}} \delta^{\;\bar{\imath}_
  {4}}_{i_{1}}\;
  \mathcal{A}^{(L)}(1_{q}, 2_{\bar{q}}, 3_{Q}, 4_{\bar{Q}}) \,,
\end{split}
\end{align}
where $S_n/Z_n$ denotes all non-cyclic permutations of $n$ indices, and
$L$ corresponds to the number of loops of the amplitudes.
In the five-point case we write
\begin{align}
  \begin{split}
  \label{eq:bareAmp}
  &\mathcal{M}^{(L)}(1_g, 2_g, 3_g, 4_g, 5_g) \big\vert_{\textrm{leading color}} = 
  N_c^L \sum_{\sigma\in S_5/Z_5} \mathrm{Tr}\left(
  T^{a_{\sigma(1)}} \ldots
  T^{a_{\sigma(5)}} \right)\\
  &\qquad\qquad\qquad\qquad\qquad\qquad
  \times \mathcal{A}^{(L)}({\sigma(1)}_g, {\sigma(2)}_g, {\sigma(3)}_g, 
  {\sigma(4)}_g, {\sigma(5)}_g)\,, 
  \end{split}\\
  \begin{split}
  \label{eq:ColorDec2Q}
  &\mathcal{M}^{(L)}(1_q, 2_{\bar{q}}, 3_g, 4_g, 5_g) \big\vert_{\textrm{leading color}} 
    =  N_c^L \sum_{\sigma\in S_3} 
  \left( T^{a_{\sigma(3)}} T^{a_{\sigma(4)}} T^{a_{\sigma(5)}} \right)^{\;\bar{\imath}_2}_{i_1} \\
  & \qquad\qquad\qquad\qquad\qquad\qquad
  \times \mathcal{A}^{(L)}(1_q,2_{\bar{q}},\sigma(3)_g,\sigma(4)_g,\sigma(5)_g)\,,
  \end{split}\\
  \label{eq:ColorDec4Q}
  &\mathcal{M}^{(L)}(1_q, 2_{\bar{q}}, 3_Q, 4_{\bar{Q}}, 5_g)
  \big\vert_{\textrm{leading color}} 
  = 
  N_c^L \,(T^{a_5})^{\;\bar{\imath}_{2}}_{i_{3}} \delta^{\;\bar{\imath}_{4}}_{i_{1}}\;
  \mathcal{A}^{(L)}(1_{q}, 2_{\bar{q}}, 5_g, 3_{Q}, 4_{\bar{Q}}) \, \nonumber\\
 &\qquad\qquad\qquad\qquad\qquad\qquad
+ N_c^L\,(T^{a_5})^{\;\bar{\imath}_{4}}_{i_{1}} \delta^{\;\bar{\imath}_{2}}_{i_{3}}\;
\mathcal{A}^{(L)}(1_{q},2_{\bar{q}},3_{Q},4_{\bar{Q}},5_g) \,.
\end{align}

\subsection{External helicity states}
\label{app:wavefunctions}
In this section we collect the conventions used in \Caravel{}
for the spinors and polarization states.
For floating-point computations, the spinors must
be constructed so that they are numerically stable, which means
different conventions are chosen depending on the phase-space point.
As an example, if $p_+$ is not small, we take
\begin{equation}
\begin{split}
 u_+(p) &= v_-(p) = \ket{p} = \frac{\sqrt{|p_+|}}{p_+}
 \begin{pmatrix} p_+ \\ p^\perp_+\end{pmatrix}\;, \\
 u_-(p) &= v_+(p) = |p] = \frac{1}{\sqrt{|p_+|}}
 \begin{pmatrix} p_+ \\ p^\perp_- \end{pmatrix}\;,
\end{split}
\end{equation}
where we defined
\begin{equation}
 p_+ = p^0 + p^3\;, \qquad p^\perp_+ = p^1 + ip^2\;, \qquad p^\perp_- = p^1 - ip^2\;.
\end{equation}
We will often use the particle index instead of its momentum to denote
the spinors. That is, for a particle with index $i$ of momentum $p_i$
we will write $\ket{i}$ and $|i]$.

For massless external vector bosons with four-momentum $p$, the polarization 
states are defined in terms of a light-like reference vector $n^\mu$ as
\begin{equation}
 \epsilon^\mu_+(p,n) = \frac{ \bra{n|\bar{\sigma}^\mu}p]}{\sqrt{2}\braket{n|p}}\;, \qquad
 \epsilon^\mu_-(p,n) = -\frac{ [n\ket{\sigma^\mu|p}}{\sqrt{2}[n|p]}\;,
\end{equation}
where $\sigma^\mu = (1,\sigma^i)$, $\bar{\sigma}^\mu = (1, -\sigma^i)$ and $\sigma^i$ are the 
Pauli matrices.
Amplitudes are independent of the specific choice of auxiliary vector $n$.
Finally, for the computation of gravity amplitudes, the external graviton 
states are constructed from the polarization vectors given above. The two
transverse polarization states read
\begin{equation}
 h^{\mu\nu}_{--}(p,n) = \epsilon^\mu_-(p,n)\epsilon^\nu_-(p,n)\;, \qquad
 h^{\mu\nu}_{++}(p,n) = \epsilon^\mu_+(p,n)\epsilon^\nu_+(p,n)\;,
\end{equation}
where as before $p$ is the four momentum of the graviton and $n$ is a massless
auxilliary vector.

In the case of finite-field evaluations, we would like the external helicity states to be rational
functions of the momentum components.
To achieve this, we exploit the fact that Weyl spinors are defined up to a little group scaling
\begin{equation}
 |p\rangle \to z_p~|p\rangle\;, \qquad |p] \to \frac{1}{z_p}~|p]\;.
\end{equation}
With $z_p=\sqrt{p_+}$, we obtain 
\begin{equation}
\begin{split}
 u_+(p) &= v_-(p) = \ket{p} = \begin{pmatrix} p^0 + p^3 \\ p^1 - p^2 \end{pmatrix}\;, \\
 u_-(p) &= v_+(p) = |p] = \frac{1}{p^0+p^3}\begin{pmatrix} p^0 + p^3 \\ p^1+p^2 \end{pmatrix}\;,
\end{split}
\end{equation}
where the components are given for the alternating signature $(+,-,+,-)$, in which the spinors can 
be rendered explicitly real. This choice of metric does not affect the value of Lorentz-invariant quantities,
such as the normalized amplitudes $A^{(L)}$ defined in \cref{eq:swNorma} below.

\subsection{Spinor weights of helicity amplitudes}
\label{sec:spinor_weight}

In this section we present our conventions for a spinor-weight normalization
which allows one to construct Lorentz-invariant objects
from helicity amplitudes in \Caravel. 
We use a generalization of the normalization factor introduced in
ref.~\cite{Badger:2016uuq} with the conventions specified in the previous section.

For an $n$-point amplitude,
let $\mathcal{C} = \{h_1,\,\ldots{},\,h_n\}$ be the sequence of the helicity states of 
$n_v$ vector bosons and $n_f$ fermion pairs ($n = n_v + 2 n_f$),
labelled by their particle index.
We then construct the spinor weight $\Phi_\mathcal{C}$ associated to $\mathcal{C}$ as
\begin{align}
  \Phi_\mathcal{C} = \prod_{i=1}^{n_v}\mathcal{\omega}^{\mathrm{sign}(h_i)}_{\mathsf{v}_i} ~ \prod_{i=1}^{n_f} \eta_{\mathsf{f}^-_i \mathsf{f}^+_i}.
\end{align}
Here, $\mathsf{v}$ and $\mathsf{f}^\pm$ are the (order-preserving) subsequences 
of the index sequence of $\mathcal{C}$
corresponding to vector-boson states and fermion states with $h_i = \pm\frac{1}{2}$ respectively.\footnote{
Note that this pairs fermion states based strictly on the ordering of 
$\mathcal{C}$ and disregards any other
quantum numbers of the corresponding particles. This might not be what is intuitively anticipated.
}
The weights $\omega^{\pm}_i$ and $\eta_{ij}$ are given by
\def\spaa#1#2{\langle #1 #2 \rangle}
\def\spab#1#2#3{\left\langle #1 \middle\vert #2 \middle\vert #3 \right]}
\def\spbb#1#2{\left[#1 #2\right]}
\begin{align}\begin{split}
    \omega_1^+ &=  \frac{\spbb{1}{2}\spaa{3}{2}}{\spaa{1}{3} }, \qquad
    \omega_i^+ =  \frac{\spaa{1}{3}}{\spaa{i}{1}^2 \spbb{1}{2} \spaa{3}{2} } \quad \text{for} \quad i \geq 2, \\
    \omega_i^- &= \frac{1}{\omega^+_i}, \qquad
    \eta_{i j} = \spaa{i}{k_{ij}}\spbb{k_{i j}}{j},
\end{split}\end{align}
where $k_{i j}$ is the smallest positive integer such that $k_{ij}\neq i$ and $k_{ij}\neq j$.
The spinor weights for the graviton helicity states are $\left(\omega^{\pm}_i\right)^2$.
We can then write the amplitudes as
\begin{equation}\label{eq:swNorma}
  \mathcal{A}^{(L)}(1_{p_1}^{h_1},\ldots,n_{p_n}^{h_n})=
  \Phi_\mathcal{C}\,A^{(L)}(1_{p_1}^{h_1},\ldots,n_{p_n}^{h_n})
\end{equation}
where $A^{(L)}(1_{p_1}^{h_1},\ldots,n_{p_n}^{h_n})$ is Lorentz invariant.

\section{Momentum-twistor parametrizations}
\label{app:twistor}
Here we list explicitly the momentum-twistor parametrization~\cite{Hodges:2009hk} 
used to rationalize the external on-shell momenta. We provide explicit
relations with the  Mandelstam invariants $s_{ij} = (p_i + p_j)^2$,
where $p_i$ denotes the momentum of the external particle with index $i$.
We recall that the conventions in \Caravel{} are that all external momenta are
outgoing.

\subsection*{Four-point kinematics}
In order to construct four-point rational momenta we use the momentum-twistor
parametrization of ref.~\cite{Badger:2013gxa}, which is given by
\begin{equation}
 Z = \begin{pmatrix} |1\rangle & |2\rangle & |3\rangle & |4\rangle \\ 
|\mu_1] & |\mu_2] & |\mu_3] & |\mu_4] \end{pmatrix} = \begin{pmatrix}
1 & 0 & -\frac{1}{s_{12}}& -\frac{1}{s_{12}} - \frac{1}{s_{23}} \\
0 & 1 & 1 & 1 \\
0 & 0 & 1 & 0 \\
0 & 0 & 0 & 1
\end{pmatrix}\;.
\end{equation}
With this parametrization the resulting momenta are rational in $s_{12}$ and
$s_{23}$.
\subsection*{Five-point kinematics}
For five-point kinematics we use the parametrization given in
refs.~\cite{Peraro:2016wsq,Badger:2017jhb}
\begin{equation}
Z = \begin{pmatrix} |1\rangle \!& |2\rangle \!& |3\rangle \!& |4\rangle \!& |5\rangle \\
|\mu_1] \!& |\mu_2] \!& |\mu_3] \!& |\mu_4] \!& |\mu_5]
\end{pmatrix}=
\begin{pmatrix}
1 & 0 & \frac{1}{x_4} & \frac{1+x_0}{x_0x_4} & \frac{1+x_1(1+x_0)}{x_0x_1x_4}\\
0 & 1 & 1 & 1 & 1\\
0 & 0 & 0 & \frac{x_2}{x_0} & 1\\
0 & 0 & 1 & 1 & \frac{x_2 \!-\! x_3}{x_2}
\end{pmatrix}.
\end{equation}
The Mandelstam variables are then parametrized by
\begin{align}\begin{split}
 s_{12} &= x_4\;, \\
 s_{23} &= x_2x_4\;, \\
 s_{34} &= x_4\left[ \frac{(1+x_1)x_2}{x_0} + x_1(x_3-1)\right]\;, \\
 s_{45} &= x_3x_4\;, \\
 s_{51} &= x_1x_4(x_0-x_2+x_3)\;, \\
 \mathrm{tr}_5 &= i\epsilon(p_1,p_2,p_3,p_4)  \\
 &= x_4^2\left[ x_2(1+2x_1)+x_0x_1(x_3-1) - \frac{x_2(1+x_2)(x_2-x_3)}{x_0}\right]\;.
\end{split}\end{align}

%
%


\bibliographystyle{elsarticle-num}
\bibliography{caravel}







\end{document}